\documentclass[pra,aps,twocolumn]{revtex4}
\usepackage{amsfonts}
\usepackage{amssymb}
\usepackage[dvips]{graphicx}

\usepackage[T1]{fontenc}
\usepackage[latin2]{inputenc}

\usepackage{amsmath}
\usepackage{bbm}

\def\duzomniejsze{<\kern-.7mm<}
\def\duzowieksze{>\kern-.7mm>}

\def\textbf#1{{\bf #1}}
\def\beq{\begin{equation}}
\def\eeq{\end{equation}}
\def\be{\begin{equation}}
\def\ee{\end{equation}}
\def\ben{\begin{eqnarray}}
\def\een{\end{eqnarray}}
\def\beqa{\begin{eqnarray}}
\def\eeqa{\end{eqnarray}}
\def\eea{\end{array}}
\def\bea{\begin{array}}
\newcommand{\bei}{\begin{itemize}}
\newcommand{\eei}{\end{itemize}}
\newcommand{\bee}{\begin{enumerate}}
\newcommand{\eee}{\end{enumerate}}

\def\acal{{\cal A}}
\def\ncal{{\cal N}}

\def\ucal{{\cal U}}

\def\>{\rangle}
\def\<{\langle}

\def\ot{\otimes}

\def\dt#1{{{\kern -.0mm\rm d}}#1\,}

\def\sigalpe{{\sigma_\alpha'}^{\kern-.7mm E}}
\def\sigalpb{{\sigma_\alpha'}^{\kern-.7mm B}}

\newtheorem{theorem}{Theorem}
\newtheorem{proposition}{Proposition}
\newtheorem{definition}{Definition}

\def\bep{\begin{proposition}}
\def\eep{\end{proposition}}
\def\bet{\begin{theorem}}
\def\eet{\end{theorem}}
\def\bed{\begin{definition}}
\def\eed{\end{definition}}

\begin{document}


\title{Can one build a quantum hard drive? 
A no-go theorem \\ for storing quantum 
information in equilibrium systems}

\author{Robert Alicki and Micha\l\ Horodecki}

\affiliation{Institute of Theoretical Physics and Astrophysics, University
of Gda\'nsk,  Wita Stwosza 57, PL 80-952 Gda\'nsk, Poland}

\date{\today}

\begin{abstract}
We prove a no-go theorem for storing quantum information in 
equilibrium systems. Namely, quantum information cannot 
be stored in a system with time-independent Hamiltonian interacting with 
heat bath of temperature $T>0$  during time that grows with 
the number of used qubits. We prove it by showing, 
that storing quantum information for macroscopic time would imply existence 
of perpetuum mobile of the second kind. 
The general results are illustrated by the Kitaev model of quantum memory.
In contrast, classical information can be 
stored in equilibrium states for arbitrary long times. 
We show how it is possible via phase-transition type phenomena.

Our result shows that there is a fundamental difference  between quantum and classical 
information in {\it physical} terms. 
\end{abstract}

\maketitle

\section{Introduction}
Quantum computing needs at least quantum memory. Quantum memory is based 
on encoding of (say 1-qubit)
states into metastable states of a larger system.
The states are expected to survive the interaction with a heat bath of 
the temperature $T>0$. 
One requires  that the life-time of the memory grows with the number $N$
of subsystems. One can consider  two methods:
time-dependent 
\cite{DoritFT,ZurekFT,KitajewPreskill-FTmem,dorit-phase}, time independent 
"self-correcting" systems  (cf. \cite{KitaevFT}). The first method is non-equilibrium one, 
where energy is constantly dissipated and moreover the 
life-time $\tau $ satisfies $N\sim poly (\log \tau)$ which means that 
the memory is claimed to be exponentially stable. The results involve a kind of 
phase transition (cf. \cite{dorit-phase,KitajewPreskill-FTmem}), and are mainly formulated 
within  phenomenological approach (see however 
\cite{AHHH2001,TerhalB-FT,Alicki-TB,AharonovKP-FT-NM}).
The second method  uses equilibrium states. 
  

In this paper we consider time-independent method 
(time dependent case will be considered in future). We show that this method 
is not useful, i.e. quantum memory cannot be stored in equilibrium states. Put it into 
different way: one cannot build a "quantum hard drive" - device that would store 
quantum information without dissipation of energy constant in time. Conversely, 
we show how classical information can be stored in equilibrium states
(this happens for all sorts of memory devices) by means of phase transition 
type phenomena.  

To prove our no-go result use theorems from mathematical physics concerning the notion of 
passivity  (inability of drawing work by external cyclic forces) and its stronger version
complete passivity \cite{Pusz-Woronowicz78}. We  invoke  the highly nontrivial  
result, stating  that  set of 
completely passive states constitute simplex, so that no quantum superpositions can
be encoded into such states. Thus our argument does not rely on a 
particular model. We refer to The Second Law, so that the result works  
in all the situations in which the law remains valid. 
This differs our results from the issues concerning decoherence of  
Schroedinger cat states (see e.g. \cite{Davies78,Joos-Zeh,Zurek81}).

Our results imply that in equilibrium one can store quantum 
information only over microscopic periods of time. Note however,
that these microscopic time scale may  vary
from system to system, depending on the type of environment to 
which the system couples and the coupling strength, and for some systems 
may be pretty long.



\section {Classical memory and phase transitions}
\label{sec:CW}
We will show on a simple model, how classical information can be efficiently 
stored.  We will use mean-field Curie-Weiss model of ferromagnetic. Though 
unphysical, it reflects very well all the essential features 
that lead to efficient storage of classical information. 
It is well known that this model exhibits phase transition. 
We will show now that the phase transition gives rise to exponentially stable 
states. We consider $N$ spins interacting by means of Hamiltonian
\be
H=-J N X^2
\ee
where 
\be
X={1\over N} \sum_{j} \sigma_i
\ee
where $\sigma_j$ is Pauli matrix $\sigma_z$ acting on $j-th$ spin,
and summation is taken over all pairs of sites. Thus, due to interaction,
the energetically favourable configurations are when spins point the same direction.
For a fixed value of the mean magnetization $X$, denoted by $x$, the energy  amounts to 
\be
E(x)=- J N x^2.
\ee
We see that there is energy barrier between the two configuration minimizing the energy. 

However, not only the height of 
the barrier is relevant, but also the number of microstates 
for a fixed macrostate $\rho_x$.
Crossing of the barrier may be likely, if the number of states on the top of 
it is large enough. The quantity that takes into account both energy difference 
and number of states is a free energy. 
Let us consider it in more detail.  Basic processes induced by noise relevant 
for our problem are flips of single spin. Single spin-flip causes flip 
between neighbouring $x$'s changing 
$x \to x \pm {1\over N}$. Thus the whole process is a random walk on $x$ line. 
The transition probabilities between macrostates $\rho_x$ are determined 
by transition probabilities between microstates. To avoid combinatorial 
issues, we adopt the heuristic rule 
\be
{p(x\to x') \over p(x'\to x)} \simeq e^{-(E(x')-E(x))/k_BT} {\ncal(x')\over \ncal(x)} 
\ee
where $\ncal(x)=\exp[Nh({x+1\over 2})]$ denotes number of microstates, 
$h(x)=-x\ln x -(1-x)\ln (1-x)$. Since the free energy is given just by
$F(x,T)=E(x)- TS(x)$ with $S(x)= N k_B h(x)$ we obtain 
\be
{p(x\to x') \over p(x'\to x)}\simeq e^{-(F(x',T)-F(x,T))/k_BT}
\label{eq:prob-free}
\ee
The terms $E$ and $TS$ in free energy have here clear interpretation:
the first one gives rise to the Boltzmann factor, while the second one 
reports the contribution of number of states to transition probability.
Thus it is now clear that it is the shape of free energy rather than 
energy itself, that will determine stability. To see it more in detail, 
we consider $x$ as continuous variable. In such limit, the problem 
is equivalent  to Brownian motion in the potential $V$ given by free energy
(see \cite{Gardiner}).

\begin{figure}
\vskip0.5cm
\includegraphics[scale=0.5]{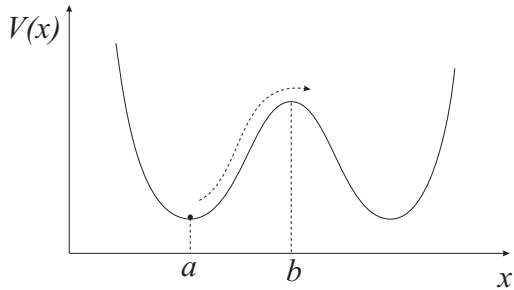}
\caption[brown]{Brownian particle in double potential well.}
\label{fig:brown}
\end{figure}

We can thus apply Kramers formula for mean exit time from a potential well (see 
fig. \ref{fig:brown})

\be
t(a\to b) \simeq {2\pi k_b T \over D} |V''(a)V(''(b)|^{-{1\over 2}} 
e^{(V(b)-V(a))/k_BT}
\label{eq:Cramers}
\ee 
where $D$ is diffusion constant \cite{Gardiner}. Thus, in our case 
we obtain,  that probability of flip is exponential 
in the difference of free energy between $x_{eq}$ 
and $x=0$ (the top of energy barrier). Thus we have to examine 
the shape of free energy, explicitly given by 
\be
F(x,T)=N(-kT h({x+1\over 2}) - Jx^2)
\ee
The function has two minima for $T<T_c$ and one minimum for $T>T_c$
(where $T_c$ is critical temperature). 
When the entropic terms dominates, the density of final states 
is so large, that the Boltzmann factor cannot prevent from 
flip between the two states, and we have single minimum.
There is only one stable state, with $x=0$ (note that this is 
the state of the largest energy).

If the energy part dominates, then the probability of flipping is exponentially 
suppressed, and we have two minima divided by barrier proportional to $N$.
Thus we have two stable phases, and according to (\ref{eq:Cramers}) probability of flipping 
between them is exponentially decreasing with $N$.

\medskip
The presented above mechanism is quite universal. Classical information can be 
efficiently stored in metastable states corresponding to local minima of the free
energy separated by barriers
which grow with the size of the system (e.g. large molecules, mesoscopic systems). 
For example the structure of protein molecules is determined by analysing 
free-energy landscape \cite{Liwo-md,Liwo2}. 


\section{Equilibrium states and passivity}
A spatially confined quantum system weakly interacting with a heat bath at the 
temperature $T$ tends to the unique equilibrium state represented by the 
Gibbs density matrix 
\begin{equation}
\rho_{\beta} = \frac{e^{-\beta H}}{{\rm Tr} e^{-\beta H}}
\label{Gibbs}
\end{equation}
where $H$ is a Hamiltonian of the system and $\beta = 1/kT$ is the inverse temperature. 
This  behavior illustrates the Zeroth Law of Thermodynamics
and can be rigorously proved for simple but generic models of reservoirs both in 
the weak coupling limit 
\cite{Davies1974-WCL} and for finite, but small enough coupling constant \cite{Der2003}. The only 
essential assumption is a kind of irreducibility of the coupling between the system and 
the bath which eliminates the appearance of conserved observables impairing equilibration 
process. This condition can be seen as the absence of decoherence free subspaces and 
subsystems  in the finite system.
\par
We are interested in the relaxation processes in the case of a system composed of 
many subsystems  (say $N$ qubits). For noninteracting subsystems and local coupling 
to the bath we observe  individual and independent relaxation processes which 
implies the thermalization time independent of the number $N$ of subsystem. 
On the  other hand, 
for strongly interacting subsystems collective phenomena can produce metastable 
states with life-times growing with $N$. We observed
such phenomenon for classical systems in the Section \ref{sec:CW} and discussed their 
applications to  classical information storage.

In principle, similar metastable states which become stationary, and resistant to local 
perturbations for $N\to\infty$ might be used to store quantum information as well. Therefore, 
it is crucial to analyse mathematical structure of such quantum states. They should 
satisfy the restrictions imposed by The Second Law of Thermodynamics. Let us recall 
Kelvin formulation of the law:

{\it  It is impossible to construct an engine which, operating in a cycle
will produce no other effect than the extraction of heat
from a reservoir and the performance of an equivalent amount of work.}

This can be  rephrased  in terms of {\it passivity}.
We say that a state $\rho$ is passive (with respect to the dynamics $\ucal_t$), 
when it is impossible to extract the energy from the system at a given state by 
means of a cyclic process (Second Law of Thermodynamics). A state is 
$n$-passive, if the state $\rho^{\ot n}$ with respect to product evolution 
$\ucal_t^{\ot n}$ is still passive. Finally, a state is called completely passive (CP), 
if it is $n$-passive for all $n$. Thus if a state is not CP, then one can extract 
energy from a finite amount of copies. 

Let us now recall how passivity and complete passivity  is described within 
quantum mechanics \cite{Pusz-Woronowicz78}.
The cyclic process is represented by time dependent perturbation $h(t)$ 
satisfying $h(0)=h(\tau)=0$.
Consider first a finite system described by the Hamiltonian $H$ with the 
eigenvectors $|j,\mu\>$ and  the possibly degenerated eigenvalues $\epsilon_j$
\begin{equation}
 H |j,\mu\> = \epsilon_j |j,\mu\> \ , \  \mu\in I_j\ .
\label{Ham}
\end{equation}
The energy change due to a time dependent perturbation $h(t)$ is given by 
$\Delta E = {\rm Tr}(\rho(U_{\tau}H U^{-1}_{\tau} - H))$ where $U_{\tau}= {\bf T}\exp\int_0^
{\tau} dt (H + h(t))$. A state $\rho$
is {\it passive} if $\Delta E \geq 0$ for all $h$. One easily shows 
that $\rho$ is passive if and 
only if
\bei
\item[(i)] $[H,\rho]=0$ and hence for every passive $\rho$ one can choose $|j,\mu\>$ in such a 
way that $\rho = \sum_{j,\mu} \lambda(j,\mu)|j,\mu\>\<j,\mu|$,
\item[(ii)]$ (\epsilon_j - \epsilon_k)(\lambda(j,\mu)-\lambda(k,\nu)) \leq 0$.
\eei 
Thus the state is passive if and only if  it commutes with the Hamiltonian, and there is 
no inversion of population, in the sense, that for any two energy levels, the upper 
level is not  more populated than the lower one. This still leaves freedom 
on degeneracies of Hamiltonian. Following \cite{Pusz-Woronowicz78} it is not 
hard to find   that 
if a state is not Gibbs one, then for some $n$ the state $\rho^{\ot n}$ 
will get inversion of population with respect to sum of single system 
Hamiltonians  $\sum_i H_i$. Therefore, for finite systems completely passive states 
are Gibbs states, hence
at the given temperature we have only a single completely passive state. 

For the infinite systems the situation is less trivial, we can have many 
CP states (e.g. different thermodynamical phases) which are the limits of metastable states 
(with increasing life-times) of the corresponding sequence of finite systems. 
Nevertheless, the structure of CP states is determined by the following theorem 
\footnote{The theorem implied by the following two seminal results: 
(i) for arbitrary system $(\acal, \ucal_t)$ at temperature $T>0$ the CP states are exactly  
so called KMS states (generalizations of Gibbs state to include infinite systems) 
\cite{Pusz-Woronowicz78};
(ii) for arbitrary system $(\acal, \ucal_t)$ at temperature $T>0$  the 
KMS states constitute simplex
\cite{RB97}.} 
valid for any system described by a $C^*$ algebra of observables
$\acal$ and a group of automorphisms (Hamiltonian dynamics) $\ucal_t$.
\medskip
\bet 
\label{thm:passive}
For arbitrary system $(\acal, \ucal_t)$ the set of CP states 
constitutes a simplex.  
\eet
This means that any CP state can be uniquely decomposed into extremal states of 
the set of all CP states. 

\section{No-go theorem for stable quantum memory}
Suppose, that we want to construct stable quantum memory 
by means of a quantum system consisting of $N$ subsystems with a specially designed 
Hamiltonian. We assume that the system interacts with a heat bath of the temperature $T>0$.

Suppose, now  that  by increasing $N$  we can arbitrarily increase the 
life-time of quantum memory. 
In particular, we can make the time much longer than any microscopic time scale, 
so that the violations of the Second Law due to fluctuations 
will be suppressed. Consider the set of states that can survive 
for macroscopic time. We call them metastable states. They have to be completely 
passive, as otherwise we could build {\it perpetuum mobile of the second kind}. 
Indeed, as explained in the previous section, if a system is in a non-CP state,
then  $n$ systems constitute a system in a nonpassive state, for some finite $n$. 
Putting now $k$ of such $n$-tuples into heat bath,
we obtain a system in equilibrium, from which one can draw work proportional to $k$
\footnote{Note, that since we requiring complete passivity we have to work 
in the limit of infinite $N$, as for finite $N$ 
only Gibbs state is CP.}.

Thus employing the Second Law, we obtain that the only states 
that can be used to store information for arbitrarily long time 
are CP states \footnote{In particular, this rules out systems 
with so called decoherence free subspaces \cite{Alicki-DFS,ZanardiR-DFS}.
This can be explained by the fact, that conditions to obtain 
decoherence free subspaces are pretty unstable.}. All other states must decay 
within microscopic time scales.   

However, according to Theorem \ref{thm:passive},
CP states form  a simplex, hence the set of metastable 
states possesses a fully classical representation. 
They can be treated as probability measures on a certain configuration space. 
Therefore CP states cannot be used for the faithful representation of the quantum states. 
It follows for example from the fact, that quantum bits cannot be faithfully conveyed 
by classical channel (see e.g. \cite{Popescu1994,Alber2001}). As a consequence only 
classical information can be preserved over macroscopic time scales.

\section{Example: Kitaev's model}
In the following we would like to discuss the model introduced by Kitaev
\cite{KitaevFT} which is supposed to be a 
good candidate for a self-correcting quantum memory. 

Kitaev considered $k\times k$ square lattice torus. On each edge there is a qubit 
(so that there are $n=2k^2$ qubits). For each vertex $s$ and  face  $p$ 
one defines operators 
\be
A_s=\sum_{j\in {\rm star}(s)} \sigma^x_j,\quad B_p=\sum_{j\in {\rm boundary}(s)} \sigma^z_j,
\ee
where "star" denotes edges that touch vertex $s$ and "boundary" denotes 
edges that surround face $p$.
The Hamiltonian is given by 
\be
H=-\sum_s A_s - \sum_p B_p
\ee
The operators $A_s$ and $B_p$ are dichotomic, and they all mutually commute.
We can form basis consisting of eigenvectors of those operators. One finds, that
for any basis vector, it has eigenvalue $1$ for even number of operators $A_s$ 
and also for even number of operators $B_p$. 
Thus the energy levels are given by the number of pairs of operators $A_s$ 
and $B_p$ to which there is eigenvalue $1$.  This can be interpreted as 
number of pairs of particles. For each level, there are many possible configurations 
of pairs. However for every configuration there is 4-dimensional 
degeneracy due to topology of the torus. Thus we can imagine the 
total system as tensor product of 4-dimensional space 
and the space determined by configurations of pairs of particles. 
The particles are (noninteracting) anyons: they exhibit non-standard statistics,
which manifests by inducing phase, when one anyon winds around the second one. 
It is the 2-qubit subsystem is the one that is expected to be noiseless, 
when the size of torus becomes large \cite{LloydZ2002-top}.

In this section we will consider the system from two points of view. 
First, we will treat it as infinite system, and discuss 
in the context of our general result.  
Subsequently, we shall consider finite system, and analyse interaction with 
a heat bath.  This will independently prove that in finite temperature, quantum 
information cannot be stored. 

\subsection{Infinite system picture}
Let us see how Kitaev's model fits into our theorem. 
To this end we will consider the limiting case of infinite system. 
The infinite system of spins is defined in terms of {\it quasilocal algebra}, spanned 
by observables that are tensor products of only finite amount of single site observables 
such as $\ldots\ot I \ot I \ot A_1 \ot \ldots \ot A_k \ot I\ot I\ot\ldots $ and 
closed in operator norm topology \cite{RB97}.
Now, one can  show that the limits of all ground (vacuum) states in the Kitaev model  
produce the same expectation values 
for all local and hence also quasilocal observables. This could be expected, 
as {\it good codewords} should be indistinguishable by local measurements. 
Therefore the infinite Kitaev system possesses a single ground state and therefore
there is no phase transition in this model and even a bit of classical information 
cannot be stored \footnote{The rigorous analysis of the infinite Kitaev model 
will be published elsewhere}. 

The physical meaning of this result is the following. The highly nonlocal 
observables which describe the encoded two qubits become, with increasing size 
of the system, more and more vulnerable to external perturbations and hence less 
and less accessible to meaningful measurements. We will see in the next section, 
that indeed the noise will wash out 
all the quantum information in time that for sure does not increase with 
the size of the system.

\subsection{Finite system interacting with heat bath}
Let us then consider the Kitaev system coupled to a heat bath  via local operators,
for example $\sigma_x^i$ and $\sigma_z^j$.  We will work
in Markovian approximation (weak coupling limit). 

The basic processes 
that can occur is i) creation of pair of neighboring anyons, ii) annihilation 
of such pair, and ii) move of an anyon. 
For example, when we hit a particular qubit with $\sigma_x$,
then, if the qubit belong to pair of anyons of type $B$, 
then the pair will be annihilated. If  there were no anyons at this place,
a pair will be created. If there was one anyon, it will be moved. 
The amplitude of the basic processes is determined by spectral density of the noise,
always however we have $p_c=e^{-\Delta E/k_BT} p_a $,
where $\Delta E$ is the energy needed to create a pair,
$p_c$ and $p_a$ are probabilities of creation, and annihilation of 
a pair of anyons respectively.
In this way in equilibrium state there is a constant density of anyons. 

One of the decoherence mechanisms is a random walk of the created anyon pairs 
which will be performed  until the anyons meet. 
It is clearly seen, that such processes will equilibrate the system 
of anyonic configurations. What about the topological qubits where  
the quantum information is to be stored? Consider such a process:
an anyon pair is created,  one of them winds around torus,
making a noncontractible loop, and they finally annihilate. 
Such a process performs a gate on one of the qubits (there are two kinds of 
noncontractible loops, and this gives gates on different qubits) \cite{KitaevFT}.
Thus such process corresponds to error, and if it is not suppressed,
we cannot store quantum information on topological qubits. 

We will now argue that quantum memory will be spoiled  at least with the probability 
independent of the size of the system (number of qubits), 
so that enlarging a system will not bring any improvement.

Let us concentrate for a while on a single pair that has been created. 
Without loss of generality, we can imagine, that one anyon is 
resting, while the other one is performing random walk. 

\begin{figure}
\vskip0.5cm
\includegraphics[scale=0.5]{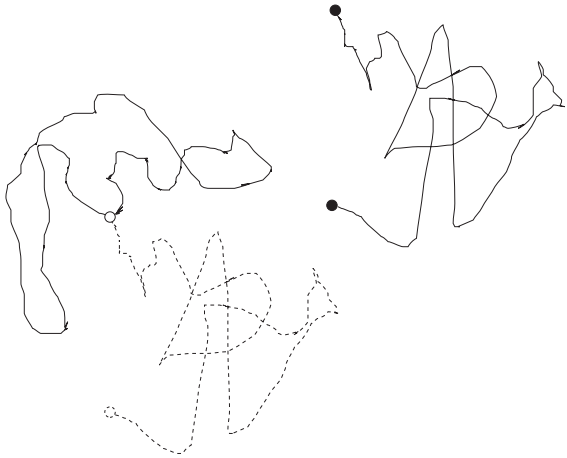}
\caption[spacer]{Random walk of anyons: the walk of one pair of anyons (white ones)
after annihilation is continued by another pair of newly created anyons (black ones).}
\label{fig:spacer}
\end{figure}

It is known that in a random walk on plane, the particle comes back 
to the origin infinitely many times. 
In our case, we do not have usual random walk, because coming back to the origin means 
annihilating the pair, and the walk is ended. If the path was short, then the process does 
not affect quantum memory.
Thus we want to estimate the probability that a long path occur. 
Since the process is Markovian (i.e. memoryless) and the average number of anyons is constant, 
we can imagine, that the walk of the just annihilated pair 
is {\it continued} by a new pair that has been just created (see Fig. \ref{fig:spacer})
In this way we have reduced the problem to the usual random walk.

It is well known,  that probability of getting away from 
a disk of the radius $L$ is proportional to $1/L^2$ and probability of a single step $p$. 
This follows e.g. from the fact that random walk is a discrete 
version of diffusion of Brownian particle.
On the other hand, the number of pairs is again proportional to $L^2$
(the proportionality constant depending on temperature).  
Thus the overall probability $p_{long}$ of making a path with a length 
of the order of a size of the system  is proportional to $p$, 
and does not depend on the size itself \footnote{One can expect that another process of
{\it fusion} of two anyonic strings into a single one will be even more efficient and yield
the decoherence rate of encoded qubits increasing with the size of the system}. 

\be
p_{long} \propto  {p\over L^2}  \times L^2
\ee

This reasoning can be performed both for $\sigma_x$ and $\sigma_z$ 
noise operators.  And existence of long paths 
will imply that both kind of nontrivial paths will be performed  with constant probability.  
Thus both topological qubits  will be subjected to random 
$X$ and $Z$ gates, which means that their state will 
tend exponentially to the maximally mixed state. 
Therefore not only the quantum information, but also the classical one will 
be washed out.  Hence also 
we will  not have phase  transition as predicted for 
infinite system with a quasi-local algebra. 
Thus we have shown, that Kitaev's system cannot provide us the means 
of storing quantum information, in accordance with our general theorem.

In conclusion, our result reveals a new, fundamental feature 
of quantum information: the latter cannot be stored in equilibrium systems. 
For practical matter, it implies, that it is impossible to construct 
quantum memories   that do not require active protection of information and 
are stable for longer  than microscopic time scales. On more fundamental level,
our result shows for the first time that there is basic difference between 
quantum and classical information not only on logical level 
(as implied e.g. by the no-cloning theorem 
\cite{Wigner-cloning,WoottersZ-cloning,Dieks-cloning}) but 
also on physical level.

\medskip

It is a pleasure to thank Pawe\l{} Horodecki and Ryszard Horodecki
for several years of numerous discussions on decoherence problems in quantum 
computers as well as for  useful discussion concerning the present work. 
R.A. is also grateful  to Daniel  Lidar, Dorit Aharonov, Michael Ben-Or and Gil Kalai
for discussions on fault tolerance. 
M.H. acknowledges discussions on the same topic with Charles Bennett, Masato Koashi, 
Debbie Leung, Noah Linden, Raymond Laflamme, John Preskill and Alexiey Kitaev.

This work is supported by Polish Ministry of Science and 
Education, grant PBZ-MIN-008/P03/2003 and EC IP SCALA.

\bibliographystyle{apsrev}
\bibliography{refmich,refpostmich,kms}





\end{document}